\documentstyle[12pt]{article}
\newcommand{\gsim}{\:\raisebox{.25ex}{$>$}\hspace*{-.75em}
      \raisebox{-.93ex}{$\sim$}\:}
\newcommand{\lsim}{\:\raisebox{.25ex}{$<$}\hspace*{-.75em}
      \raisebox{-.93ex}{$\sim$}\:}

\newcommand{\be}{\begin{equation}}
\newcommand{\ee}{\end{equation}}

\begin{document}

\begin{center}
{\bf DYNAMICS  OF THE QCD STRING WITH LIGHT AND HEAVY QUARKS}
\vspace{5mm}

{\bf E.L.Gubankova, A.Yu.Dubin}
\end{center}

\vspace{5mm}

\begin{center}
Institute of Theoretical and Experimental Physics,\\
B.Cheremushkinskaya 25, Moscow, 117259, Russia
\end{center}

\vspace{10mm}

\vspace{1cm}
\begin{abstract}

The generalization of the effective action [1] of the quark--antiquark
system  in the confining  vacuum is performed for the case of arbitrary
quark masses. The interaction of quarks is described by the averaged Wilson
loop for which we use the minimal area law asymptotics.

The system is quantized by the path integral method and the quantum
Hamiltonian is obtained. It contains not only quark degrees of freedom but
also the string energy density.

As well as in the equal masses case [1] two dynamical regimes are found
[2]: for
large orbital excitations ($l \gg 1$) the system is represented as rotating
string, which leads to asymptotically linear Regge trajectories, while at
small $l$ one obtains a potential-like relativistic or nonrelativistic
regime.

In the limiting cases of light-light and heavy-light mesons a unified
description is developed [2]. For the   Regge trajectories one obtains nearly
straight-line patterns with the slope very  close to $1/2 \pi \sigma$ and
$1/ \pi\sigma$ correspondingly. The upper bound on the light quark(s)
masses which doesn't change considerably this property of the trajectories
is also found.
\end{abstract}
\vspace{2cm}


\section
{\bf I n t r o d u c t i o n}
\vspace{5mm}

{}~~~~~ Recently the new approach to study nonperturbative large distance
dynamics of quark-antiquark system in the confining vacuum has been developed
and the Hamiltonian for the case of equal quark masses has been obtained [1].
In the present paper we consider the general case of arbitrary quark
masses.
The effective Hamiltonian of the system is derived (for a short report see
[2]) and properties of its spectrum are analysed.

With the help of vacuum correlator formalizm [3] we represent  gauge
invariant Green function of the $q\bar{q}$ system in a form where all
dynamics of the interaction is described by the averaged Wilson loop
operator. Starting from the QCD Lagrangian and making
use of the minimal area law
asymptotics for the Wilson loop we arrive to the string picture of $q -
\bar{q}$ interaction. The appearence of the string at large distances has
been discussed for a long time [4,5] (for recent calculations of the
colour-tube or string dynamics see ref. [5] ) and here the world sheet of
the string coincides with the surface  appearing in the area law of Wilson
loop.

According to the vacuum correlators method [6, 3] it is the minimal surface
which
enters the area law asymptotics. This  aproximation provides a good
calculation scheme at least for the consideration of the leading Regge
trajectories.

As in [3,1] the minimal surface is approximated as the world sheet of the
straight line connecting positions of  quark and antiquark with the
same time in the meson rest frame. We discuss the dynamical motivations of
this ansatz for the area, which leads to a  local in time string interaction
usually postulated in the flux tube models picture [5]. In this
approximation the string (minimal string) may rotate and oscillate
longitudinally (stretching and expanding).

We simplify our problem by disregarding effects due to the quark spins and
additional quark pairs creation having in mind to come back to them in
subsequent
papers and perform path integral quantization of the system
of  quark-antiquark with arbitrary masses
connected by the minimal string.

It turns out that our effective action can be represeted in a form
postulated in accordance with the flux tube picture in ref.[5]
without direct connection to the QCD Lagrangian.

We generalize the procedure for determination of the center of masses
coordinate as compared with that of the equal masses case [1] where the
symmetry
between quarks simplifies the consideration. Due to the string contribution
to the kinetic part of the action, we are to exploit the condition, that the
center of masses coordinate is decoupled from the relative one.

As well as in the equal masses case we find two dynamical
regimes of the "minimal" QCD
string with quarks of arbitrary masses. They are distinguished by the energy-
momentum distribution of nonperturbative gluonic fields along the string:
for large orbital  momenta $l \gg n_r$ the main part of the energy
(orbital momentum) is contained in the rotating
string and the resulting spectrum in the leading approximation coincides
with that of the pure string. For low values of
$l ~(l \ll n_r)$ the dynamics is described
by a relativistic or nonrelativistic (depending on quark masses)
potential-like approach with almost inert contribution from the string.
The transition between these two regimes is very
smooth and the relativization of heavy quark(s) due to increase of $l, n_r$
is briefly discussed.

For two limiting cases of heavy-light and light-light mesons a unified
description is developed. The resulting spectra in both cases are very
simple: practically linear leading Regge-trajectories and nearly parallel to
them daughter trajectories, corresponding to radial excitations with the
slopes $1/2\pi\sigma$ and $1/ \pi\sigma$ correspondingly. We estimate also
the upper bound on the light quark(s) masses, which don't change
considerably the slope of the trajectories.

We note that this pattern for the trajectories is very close to one obtained
by numerical quantization [5] of the same system which was performed without
qualitative consideration of the underlieing regimes of quark-string
interaction.

The modification of heavy quarkonia Hamiltonian due to the string
contribution to the orbital momentum is considered. The corresponding
correction to the
energy is estimated and we find the conditions under which it is comparable
with the first relativistic correction.

The paper is organized in the following way. In Section 2 we derive the
effective action of the $q \bar q$ system at large distances and discuss
the basic  approximations, which enable one to reduce initial field theory
problem to the local
three demensional quantum mechanics of quarks connected
by the straight line string. In Section 3 we represent this action
in a gaussian with respect to the quark coordinates form. The Hamiltonian
for the "minimal" QCD string with quarks is obtained in Section 4 and in
Section 5 we analyse the properties of its spectrum.

In Apperdix A intermediate steps of the transformation to the gaussian
representation are considered. Technical details for calculation of the
first corrections to zero order pure potential and pure string approximations
for the dynamics are collected in Appendixes B and C correspondingly.


\section
{\bf Ef\/fective QCD action ~of ~the $q \bar{q}$ system ~at ~large distances}
\vspace{3mm}

We start with the Green function for spinless $q\bar{q}$-system in quenched
approximation, which
in the Feynman-Schwinger representation has the following form [3]

\begin{equation}
G(x \bar{x} \mid ~y, \bar{y}) = \int^{\infty}_{0} ~ds ~\int^{\infty}_{0}
{}~d\bar{s}e^{-K-\bar{K}} ~DzD\bar{z} < W(C) >_{A}~~,
\end{equation}
where $W(C)$ is the usual Wilson loop operator

\begin{equation}
W(C) = tr ~P ~exp ~[ig ~\int_{C} ~A_{\mu} dz{\mu}]
\end{equation}
with

\begin{equation}
K = m^2_1~s + \frac{1}{4} \int^{s}_0 \dot{z}^2_{\mu} (t) dt~~,
{}~~~\dot{z}_{\mu} = \frac{d\dot{z}_{\mu}(t)}{dt}
\end{equation}
and the analogous for $\bar{K}$.

The closed contour $C$ consists of initial and final pieces $[x, \bar{x}]~,
{}~~[y, \bar y]$ ~entering the boundary conditions
and paths $z(t)~, ~~\bar{z} (\bar{t})$ of the quark and
antiquark.

At large distances $R > T_g \sim 0.2 - 0.3 ~fm$ (where $T_g$ is the vacuum
gluon correlation length) one gets from the cumulant expansion for Wilson
loop asymptotics (omitting perimeter-type terms always coming from the
quark-mass
renormalization ) [6,3]

\begin{equation}
< W(C) > ~ \approx ~exp (- \sigma S)
\end{equation}
where $S$ is the area of the  minimal surface inside the contour $C$.

This asymptotics is well confirmed by lattice simulations and allows one
to take advantage of the approximate reduction of gluodynamics to the
formation of buozeen string between quarks. As a consequence we are left
at this step with the action depending only on quark coordinates
$z_{\mu}, \bar z_{\mu}$ (since the minimal surface is determined by the
form of these trajectories).

The reparametrization from proper times $t, \bar{t}$ to the Euclidean times
$z_0
\equiv \tau~, ~~\bar{z} \equiv \bar{\tau}$ performed as in [1] yields

\begin{equation}
dt = \frac{dz_0} {\dot{z}_0}~, ~~~~~~~~~~d\bar{t} =
\frac{d\bar{z}_0}{\dot{\bar{z}}_0}
\end{equation}
which amounts [1] to the
following substitution in eq.(3) for kinetic terms

\begin{equation}
K \to K' = \int\limits^{T}_{0} \frac{d\tau}{2} [\frac{m^{2}_{1}}{\mu_1(\tau)}
+\mu_{1}(\tau) \{ 1+\dot{\vec{z}} ^2(\tau) \}] ~ ,
\end{equation}
$$ds Dz_0 \to D\mu_{1} $$
and the analogous for $\bar{K}$. Here we have introduced path integration
over the new functions, playing, as we will see, the role of dynamical quark
masses
\begin{equation}
\mu_{1}(\tau)=\frac{T}{2s}\dot{z}_{0}(\tau)~,~~~~~~
\mu_{2}(\tau)=\frac{T}{2\bar{s}} \dot{\bar{z}}_0(\tau)
\end{equation}
and $T = \frac{1}{2}(x_0 + \bar{x}_0 - y_0 - \bar{y}_0)$.

Let us discuss the approximate reduction, which we apply to the dynamics
of zero components $z_0~,~ \bar{z}_0$. Initial path integral representation
takes into account all trajectories, including those with the backward
motion in time $d \tau = \dot{z}_0 (\gamma)~d\gamma~ < ~0 $~, where the
signs of $\dot{z}_0$ ,  $\dot{\bar{z}}_0$ are negative. In the Fock space
the backward in time pieces of trajectories are responsible for the creation of
additional $q \bar{q}$--pair. Here we neglect
this backward motion of $q, \bar{q}$
trajectories since the backtracking in time quark is dragging with itself also
the
string and this enlarges the action due to the formation of foldings on the
string world sheet.

With respect to the space--time picture of the evolution it means that we
don't take into account breaking of our string into several ones due to
 the quark pairs creation from the vacuum (in addition to the neglect of
such breaking due to $q \bar q$ pairs from quark determinant
which corresponds to the quenched approximation (1) we started from).

In what follows this  no-backtracking time approximation leads to reduction
of quark dynamics to that of the valence quark (connected by the string)
Fock sector, where the conditions

\be
\mu_1(\tau)>0~,\qquad \mu_2(\tau)>0
\ee
are valid and the transformations (5) are unique.

Now one is to f\/ind reasonable ansantz for the minimal surface of
Wilson loop (4) in terms of $z$ and $\bar z$. As in [1,2] we introduce standard
compact notations
\begin{equation}
\xi  \equiv \{ \tau, \beta\}\;\;,\;\;\;
g_{ab}(\xi) \equiv \partial_a w_{\mu}\partial_{b} w^{\mu},\;\;\; a,b =
\tau,\beta
\end{equation}
so that the area $S$ is

\begin{equation}
S =
\int d^2 \xi  \sqrt{det g} ~ ,
\end{equation}
where $w_{\mu}(\tau, \beta)$
are the coordinates of the string world surface, and $$ \dot{w}_{\mu}=
\frac{\partial w_{\mu}}{\partial \tau} ~ , ~~~~~~ w'_{\mu}= \frac{\partial
w_{\mu}}{\partial \beta} ~ . $$

As well as in [7, 1] we can use in the valence quark sector (8)
the approximation that the minimal surface for given paths
$z_{\mu} (\tau)$, ~$\bar{z}_{\mu}(\tau)$ is determined by eq.(10) with
$w_{\mu}$  given by straight lines, connecting points $z_{\mu}(\tau)$ and
$\bar{z}_{\mu}(\tau)$ with the same $\tau$, i.e. we exploit the instantaneous
approximation to the interaction (10).
\begin{equation}
z_{\mu}=(\tau~, \vec{z}) ~, ~~~~~ \bar{z}_{\mu}
(\tau) = (\tau~,~\vec{\bar{z}}) ~ ,
\end{equation}
and
\begin{equation}
 w_{\mu}(\tau,\beta) = z_{\mu}(\tau)\cdot
 \beta + \bar{z}_{\mu}(\tau )(1-\beta)\;\; ,
 \;\;\;\;\;0 \preceq \beta \preceq 1 ~ .
\end{equation}
We note that this approximation to the surface geometry is inspired by two
limiting cases which are of special interest below: in the case $l=0$ one
can make use of the f\/lat dynamics of quarks and in the  limit $l\rightarrow
\infty$ quarks and  the  string are moving along typical trajectories of
the double helycoid type [7], for which the minimal area indeed is formed by
the
straight-lines. It amounts ef\/fectively to the elimination of the second
time from the problem and corresponds to the instantaneous formation of the
string in accordance with a position of quarks.

Combining the results we obtain the total effective action in the form

\be
   A \equiv K' + \bar{K}' + \sigma \int\limits^T_0
   d \tau \int\limits^1_0 d \beta \sqrt{detg} \equiv
   K' + \bar{K}' + A_{str.}
\ee
where the kinetic terms of quarks are
$$
   K' + \bar{K}' = \int\limits^T_0 d \tau \frac{1}{2}
   \left[ \left( \frac{m^2_1}{\mu_1(\tau)} +
   \mu_1(\tau) \left\{ 1+ \dot{\vec{z}}^{~2} (\tau) \right\}
   \right) + \right.   $$
\be
   \left. + \left( \frac{m^2_2}{\mu_2(\tau)} + \mu_2
   (\tau) \left\{ 1 + \dot{\bar{z}}^{~2} (\tau) \right\}
   \right) \right]
\ee
We emphasize, that the presence of the      terms $K' + \bar{K}'$
violates        the condition that the ends of the string move
with the velocity of light.

We stress [1] also, that it is the valence-quark approximation (8) together
with the equal time straight-line ansatz (12), which enables one to reduce
at large interquark disances nonlocal
four dimensional dynamics (1) to the local in time three dimensional
 dynamics (13).

After integration over $\mu_1, \mu_2$ our action (13) obtained from QCD
Lagrangian coincides with that of ref.[5] postulated in accordance with
the flux tube picture of the interaction.

But as we will see our way of formulation and solution of the Hamilonian
problem is different from numerical one used in [5]. The main advantage of
our analitical approach is the possibility to reveal underlieing
dynamical regimes of quark-string interaction discussed in Section 5.

\section
{\bf Gaussian representation for the effective action}
\vspace{3mm}

A direct procedure of quantization of eq.(13) is dif\/ficult due to the
square root term and as in [1, 2] we use the auxiliary fields approach [8]
to get rid of it.

The string part of action $A_{str.}$ in (13) is equivalent [1]
(for the straight line ansatz) to the following
one quadratic in quark coordinates[8]
\be
   A_{str.} = \int\limits^T_0 d \tau \int\limits^1_0 d \beta
   \frac{\nu}{2} \left[ \dot{w}^2 + (\sigma / \nu)^2
   w'^2 - 2 \eta (\dot{w}w') + \eta^2 w'^2 \right]
\ee
where the field $\nu$ will play the role of the string energy density and
the conditions (11) for $z_{\mu}(\tau),~ \bar{z}_{\mu}(\tau)$
are  implyed.

The auxiliary  fields $\nu (\tau, \beta) \ge 0$  and
$\eta (\tau, \beta)$  are integrated out together
with $\mu_i(\tau)$ in the full path integral representation for $G$
\be
G = \int D\vec{z} D\vec{\bar z} D\nu D \eta D \mu_1 D \mu_2 e^{-A} ~.
\ee
We emphasize that (up to the preexponential factor, which is immaterial
in our case
for the derivation of effective action) the integration over $\mu_1, \mu_2,
\nu$ and $\eta$ effectively amounts to the replacement of them
by their extremum values [1].

In the valence quark Fock sector (8), (12) one can introduce
relative $r_{\mu}$ and center of masses $R_{\mu}$ coordinates in a
selfconsistent way and fix the rest frame of the meson. Contrary to the
equal masses case [1] here one is to use explicitly the condition that
$\dot{R}_{\mu}$ is decoupled from $\dot{r}_{\mu}$.

First we note that the action (13), (15) must not  contain an explicit
dependence on $R_{\mu}$ (for the total momentum $P_{\mu}$ to be conserved).
Therefore the relative coordinate is to be defined as
\be r_{\mu} = w'_{\mu} = z_{\mu}(\tau)
- \bar{z}_{\mu}(\tau) ~.
\ee

Second, to diagonalize the quadratic in $\dot{z}~, ~\dot{\bar{z}}$
kinetic part of $A$ in such a way that the Jacobian

\be
   \frac{\partial(R, r)}{\partial(z, \bar{z})} ~ = ~ 1
\ee
we are to introduce
\be
   R_{\mu} ~=~ \zeta (\tau) z_{\mu}(\tau) ~+~ (1 - \zeta(\tau))
\bar{z}_{\mu}(\tau)
\ee
where the parameter $\zeta(\tau)$ is determined from the condition, that
$\dot{R}$
is decoupled from  $\dot{r}$.

Actually the action can be rewritten in terms of $\dot R_{\mu}, \dot r_{\mu},
r_{\mu}$  as
$$
   A = \int\limits^T_0 d \tau \left[ \frac{m^2_1}{2\mu_1} +
   \frac{m^2_2}{2\mu_2} + \frac{1}{2} \left\{ a_1 \dot{R}^2 +
   2a_2 (\dot{R}\dot{r}) - 2 c_1 (\dot{R} r) - \right. \right. $$
\be
   \left. \left. - 2 c_2 (\dot{r} r ) +
   a_3 \dot{r}^2 + a_4 r^2 \right\} \right]
\ee
where we have used $w_{\mu} = R_{\mu} + (\beta - \zeta) r_{\mu}$
to express $\dot{w}$ in terms of $\dot{R},~ \dot{r}$ and the
following notations have been  introduced
$$  a_1 = \int\limits^1_0 d \beta (\mu_1 + \mu_2 + \nu) ~, ~~~~~
    a_3 = \int\limits^1_0 d \beta \left( \mu_1  (1 - \zeta)^2  +
          \mu_2 \zeta^2 + (\beta - \zeta)^2 \nu \right) ,   $$
\be
\begin{array}{ll}
    a_2 = \int\limits^1_0 d \beta \left(\mu_1 - \zeta (\mu_1 + \mu_2)
      + (\beta - \zeta) \nu \right) , &
   a_4 =  \int\limits^1_0 d \beta \left( \frac{\sigma^2}{\nu}
     + \eta^2 \nu \right) ,\\
    c_1 =  \int\limits^1_0 d \beta \eta \nu ,  &
    c_2 =  \int\limits^1_0 d \beta \eta (\beta - \zeta) \eta \nu ~.\\
\end{array}
\ee

For the diagonalization of the kinetic part of $A$ one is to put
$a_2 = 0$, which leads to
\be
   \zeta (\tau) = \frac{\mu_1 (\tau) + < \beta > \int
   \nu(\tau, \beta) d \beta}
   {\mu_1 (\tau) + \mu_2 (\tau) + \int \nu (\tau, \beta) d \beta}
\ee
where
\be
   < \beta > = \frac{\int \beta \nu d \beta}{\int \nu d \beta} ~.
\ee

It should be stressed, that eq.(22) for $\zeta (\tau)$
which determines the $_"$center of masses" coordinate (19) involves
not only quark dynamical masses $\mu_i$ but also the dynamical
string energy density $\nu(\tau, \beta )$. This  corresponds
to the string contribution to the kinetic part of  action
(20).

Eqs.(19), (22) generalize the equal masses case $m_1 = m_2$ [1] where one
gets for the extremal values of $\mu_i$
\be
\mu_1 (\tau) ~ = ~ \mu_2 (\tau) ~.
\ee
Since the extremal values of the function
$\nu(\beta)$ (that only contribute to the action) are even under the
exchange $\beta - \frac{1}{2} \to - (\beta - \frac{1}{2})$ (due to the
symmetry under permutation of the ends of the string) one obtains from
(22), (24) for $m_1 = m_2$.
\be \int (\beta -
   1/2) \nu d \beta ~ = ~ 0 ~,
\ee
\be
   \zeta(\tau) ~ = ~ 1/2 ~,
\ee
\be
   \vec{R}(\tau) ~ = ~
   \frac{1}{2}\left( \vec{z}(\tau) + \vec{\bar{z}} (\tau) \right) ~.
\ee

In the end of this section let us f\/ix the frame of reference.
For this purpose we perform usual canonical transformation from
$\dot{\vec{R}}$ to the total momentum $\vec{P}$
\be
   \int D \vec{R} exp \left[ i \int L (\dot{\vec{R}}, ...) d
   \tau \right] = \int D \vec{R} D \vec{P} exp \left[
   i \int \left\{ \vec{P}\dot{\vec{R}} - H (\vec{P}, ...)
   \right\} d \tau \right]
\ee
where $H(\vec{P}, ...) = \vec{P}\dot{\vec{R}} - L(\dot{\vec{R}},
...)$ and conserved total momentum is def\/ined as
\be
   \vec{P} ~ = ~ \frac{\partial L (\dot{\vec{R}}, ...)}
   {\partial \dot{\vec{R}}} ~.
\ee

Meson rest frame correspons to $\vec{P} = 0 $. Taking into
account conditions (11) leading to
\be
   \dot{R}_0 (\tau) = 1 , ~~~~~~ r_0 (\tau) = 0
\ee
we obtain for the effective action
$\tilde{A},$  equal to $H (\vec{P} = 0, \ldots)$
\be
   \tilde{A} = \frac{1}{2} \int\limits^T_0 d \tau \left[
   \frac{m^2_1}{\mu_1} + \frac{m^2_2}{\mu_2} + \frac{1}{a_1}
   \left\{ a_3 a_1 \dot{r}^2 - 2 c_2 a_1 (r \dot{r}) +
   (a_4 a_1 - c^2_1) r^2 + a^2_1 \right\} \right] ~.
\ee
\vspace{8mm}

\section
{\bf Derivation of the Hamiltonian for the "minimal" QCD string with
quarks}
\vspace{5mm}

Action (31) contains (through the dependence of $a_i, c_i$)
the auxiliary functions $\mu_1~, \mu_2~, \nu~,
\eta~$. In contrast to the first three ones (playing the roles of quark
dynamical masses and string energy density respectively as one can see for
example from
eq. (22) ) the function $\eta (\tau~, \beta)$~ is an intermediate one. To
derive  the ef\/fective Hamiltonian for
                the  minimal      QCD-string with quarks we perform
gaussian integration ($\tilde{A}$ is quadratic
form in $\eta$) over $\eta$ to obtain the effective action $A'$
\be
   \int D \eta (\tau, \beta) exp [ -\tilde{A}] ~\sim~ e^{-A'} ~.
\ee
It amounts to the  substitution into $\tilde{A}$ the extremal
value of $\eta$ (see Appendix A for details)
\be
   \eta_{ext} (\tau, \beta) = \frac{(\dot{\vec{r}} \vec{r})}{\vec{r}^2}
   \left( \beta - \frac{\mu_1}{\mu_1 + \mu_2 } \right)
 \ee
which leads to the following definition of $A'$
$$
A' = \frac{1}{2} \int\limits^T_0 d \tau \left[ \frac{m^2_1}
  {\mu_1} + \frac{m^2_2}{\mu_2} + \mu_1 + \mu_2 + \int \nu d \beta
  + \left\{ \mu_1 (1 - \zeta)^2 + \mu_2 \zeta^2
   + \int (\beta -
   \zeta)^2 \nu d \beta \right\} \right. \times
$$
\be
  \left. \times ~ \frac{[\dot{\vec{r}} \times  \vec{r}]^2}{\vec{r}^2}
  + \tilde{\mu} \frac{(\dot{\vec r} \vec r)^2}{\vec r^{~2}}
  + \vec{r}^{~2} \int \frac{\sigma^2 d \beta}{\nu} \right]
\ee
where
$$  \tilde{\mu} ~ = ~ \frac{\mu_1 \mu_2}{\mu_1 + \mu_2}  ~.$$

Expression (34) forms the basis of our further calculations.

To perform canonical transformation from $\dot
{\vec{r}}$ to $\vec{p}$ we separate longitudinal and transverse
with respect to $\vec{r}$  components of $\dot{\vec{r}}$
\be
   \dot{\vec{r}}^{~2} ~ = ~ \frac{1}{\vec{r}^{~2}} \left\{ (\dot{\vec{r}}
   \vec{r}~)^2 + [\dot{\vec{r}} \times \vec{r}~]^2 \right\}
\ee
and obtain   for the logitudinal and transverse components of
the momentum  respectively
\be
   \vec{p}^{~2}_r ~ \equiv ~ \frac{(\vec{p} \vec{r}~)^2}{\vec{r}^{~2}}
   ~ = ~ \tilde{\mu}^2 \frac{(\dot{\vec{r}} \vec{r}~)^2}{\vec{r}^{~2}}
\ee
\be
    \vec{p}^{~2}_T \equiv \frac{[\vec{p} \times \vec{r}~]^2}{\vec{r}^{~2}} =
    \left( \mu_1 (1 - \zeta)^2 + \mu_2 \zeta^2 + \int\limits
    ^1_0 d \beta (\beta-\zeta)^2 \nu  \right) ^2
    \frac { [\dot{\vec{r}} \times \vec{r}~]^2}{\vec{r}^{~2}} ~.
\ee

We note that the string energy density $\nu$ doesn't contribute into the
longitudinal component of the momentum, because the momentum density
of the string
\be
   P^{str}_{\mu} = \frac{\partial \left( \sigma \sqrt{\dot{w}^2
   w'^2 - (\dot{w}w'~)^2 } \right) }{\partial \dot{w}_{\mu}
   (\tau, \beta) }
\ee
is orthogonal to $w'_{\mu} = r_{\mu}$
\be
   \left( P^{str} w' \right)  =  \left(  P^{str} r \right) = 0 ~.
\ee

The standard derivation  of $H$ from the action (13)
yields (in the Minkowski space-time) for the Hamiltonian
in functional integral
$$ H (p, r, \nu, \mu_1, \mu_2 ) = \frac{1}{2} \left[
   \frac{(\vec{p}^{~2}_r + m^2_1)}{\mu_1} + \frac{(\vec{p}^{~2}_r +
   m^2_2)}{\mu_2} + \mu_1 + \mu_2 + \vec{r}^{~2} \int\limits
   ^1_0 \frac{\sigma^2 d \beta}{\nu} +  \right. $$
\be
   \left. + \int\limits^1_0 \nu d \beta + \frac{\hat{\vec{L}}^{~2} /
   \vec{r}^{~2}}{\left(\mu_1 (1-\zeta)^2 + \mu_2 \zeta^2 +
   \int\limits^1_0 d \beta (\beta - \zeta)^2 \nu \right)} \right]
\ee
where $\hat{\vec{L}}^2 = \vec{p}^{~2}_T ~\vec{r}^{~2} $.

This is the resulting Hamiltonian for the $_"$minimal" QCD string
with quarks. For the case of equal masses $m_1 = m_2$, which
amounts to the possibility of substitutions (24) - (27), it has been derived
in ref.[1]. Approximate expressions for eq.(40)
with disregard of string contribution to the orbital momentum
were obtained in [7].

Let us stress again, that eq.(34) contains auxiliary f\/ields $~\mu_1,
{}~\mu_2, ~\nu$.
After integration over them in the full path integral representation
(16) only saddle points values $\mu^{ext}_{i} (\tau), ~\nu^{ext}
(\tau, \beta)$
\be
   \frac{\partial H}{\partial \mu_i (\tau)}
   \left|_{\mu_i = \mu_i^{ext}, ~\nu = \nu^{ext}} ~
   \right. = 0, ~~~~i = 1, 2 ~;
   ~~~~~ \frac{\delta H}{\delta \nu (\tau, \beta)}
   \left|_{\nu = \nu^{ext}, ~\mu_i = \mu_i^{ext}} ~\right. = 0
\ee
contribute [1] in the ef\/fective Hamiltonian. The conditions (41)
lead to the following equations for the extremal values of
auxiliary f\/ields
\be
   \frac{\vec{p}^{~2} + m^2_1}{\mu^2_1 (\tau)} = 1 - l(l+1)/\vec{r}^{~2}
   \left( \frac{(1 - \zeta)^2}{a^2_3} - \frac{1}{\mu^2_1} \right) ~ ,
\ee
\be
   \frac{\vec{p}^2 + m^2_2}{\mu^2_2 (\tau)} = 1 - l(l + 1)/\vec{r}^{~2}
   \left( \frac{\zeta^2}{a^2_3} - \frac{1}{\mu^2_2} \right) ~ ,
\ee
\be
   \frac{\sigma^2}{\nu^2(\tau, \beta)} \vec{r}^{~2} = 1 - l(l+1)/ \vec{r}^2
   \frac{(\beta - \zeta)^2}{a^2_3} ~ ,
\ee
where $a_3 = \mu_1 (1 - \zeta)^2 + \mu_2 \zeta^2 + \int d \beta
(\beta-\zeta)^2 \nu d \beta$ and $\zeta$ is def\/ined by eq.(22).

Only after substitution of these extremal values into the path integral
Hamiltonian (40) one is to construct (performing proper Weil ordering
[9]) the operator Hamiltonian acting on the wave functions.


\section
{\bf Dynamical regimes of the "minimal" QCD string with quarks}
\vspace{5mm}

In this section we perform     analysis of dynamics of the "minimal" QCD
string with quarks. First we consider the case when
both (heavy) quarks are nonrelativistic.
We calculate the string contribution to the orbital momentum of heavy
quarkonia and find the domain of quantum numbers $l, n_r$, where the
corresponding correction to the energy is comparable with the leading
relativistic correction.
Second we analyse another limit of one or
both light quarks ($m_i \lsim \sqrt{\sigma}$) and find two relativistic
regimes for large and small orbital momentums. Finally we discuss  finite
mass effects and the transition from  nonrelativistic dynamics of heavy
quark(s)
to the relativistic one.
\vspace{5mm}


\centerline{\bf  A.~The limit of nonrelativistic potential dynamics}
\vspace{5mm}

Let us consider the heavy quarks limit of the QCD string with quarks
\be
   m_1, ~~m_2 ~\gg  ~\sqrt{\sigma} ~.
\ee

In order to deal with a realistic case we impose also the condition that
the characteristic distance $< r >$ between heavy quarks is more than
the vacuum gluon correlation length $T_g \sim 0.2 fm$ and the dynamics is
governed by the confining potential. It amounts to the following constraint
on the masses
\be
   < r > ~ \sim ~ (2 \tilde m \sigma)^{-1/3} ~ \gsim ~ T_g ~ ,
\ee
where $\tilde m$ is the reduced quark mass.

In this case one easily gets in the leading order from eqs.(42)-(44)
$$   \mu_i^{ext} = m_i ~, ~~~~ i = 1, 2 ~ ;$$
$$   \nu^{ext} = \sigma \mid \vec{r} \mid  ~ ,$$
$$  \zeta = \frac{m_1}{m_1 + m_2} ~ ,$$
with the second term in righthand sides of eqs.(42)-(44) being  much
less than all others. Therefore we obtain neglecting all relativistic and
string corrections
the nonrelativistic Hamiltonian in the  form
(for large distances between quarks)
\be
    H =  m_1 + m_2 + \frac{1} {2\tilde{m}} \vec{p}^{~2} +
    \sigma \mid \vec{r} \mid ~,
\ee
where
$$ \tilde{m} ~ = ~ \frac{m_1 m_2}{m_1 + m_2} .$$

The first corrections to eq.(47) can be easily obtained from the
eqs.(40)-(44) with the result (for  equal masses $m_1 = m_2 = m$ case)
\be
    \Delta H^{rel} = - \frac{1}{4} \frac{p^4}{m^3} - \frac{l(l+1)
    \sigma}{6 m^2 r} ~ .
\ee
The first term comes from the expansion of the
ordinary square root $2 \sqrt{\vec p^{~2} + m^2}$ and the second one
corresponds to the string contribution to the orbital momentum (see
the next subsection for a more detailed discussion).

Simple estimates (see Appendix B for details) give the following results for
$\Delta E^{rel}$
\be
   \Delta E^{rel}_{n_r,l} = - \frac{E^{(0)^2}_{n_r,l}}{36m}
   -  \frac{l(l+1) \sigma^2}{4 m^2 E^{(0)}_{n_r,l}} ~ ,
\ee
where $E^{(0)}_{n_r,l}$  is the eigenvalue of the nonrelativistic
Hamiltonian (47) after subtraction of heavy masses. One can represent
$E^{(0)}_{n_r,l}$ in the form [7]
\be
   E^{(0)}_{n_r,l} = (2m)^{-1/3} \sigma^{2/3} a_{n_r, l} ~ ,
\ee
where $a_{n_r, l}$  is the eigenvalue of the corresponding dimensionless
Hamiltonian. Taking for example $n_r = 0, l = 1, a_{0, 1} = 3.36 $ [7]
we obtain that for not very large $l \ne 0, n_r$ the string contribution
(the second term in eqs. (48), (49) )  ~is of order of the first relativistic
correction.

Making use of the asymptotical expression [7]
\be
   a_{n_r,l}  \approx \left[ \frac{n}{2} \right]^{2/3}
   \left( 3 + O \left( \frac{2n_r}{n} \right) \right) ~ ,
   ~~~~~~~~~~~l \gg n_r ~ ,
\ee
where $n = n_r + l + 1$ we conclude that for large $l \gsim n_r$
these corrections remain to be comparable while for large $n_r \gg l$
the relativistic one becomes obviously dominant.

In conclusion we  note that the string correction to the
nonrelativistic Hamiltonian (47) was also  considered in ref.[10].
\vspace{5mm}


\centerline{\bf B. ~Relativistic potential regime of the system for small
orbital momenta}
\vspace{3mm}

Let us consider relativistic dynamics of our system and prove
the existence of two relativistic dynamical regimes of the
"minimal" QCD string with quarks: potential one for small orbital
momenta (or $n_r \gg l$) and the string one for large $l \gg n_r$,
 which join each other very smothly. For the equal masses case
$m_1 = m_2$ this analysis has been performed in ref.[1] and here
we generalize it for the system with arbitrary masses of quarks.

We begin our consideration of the general case with derivation of the
relativistic  Hamiltonian for zero orbital momentum $l = 0$. As we shall
prove it provides a good zero approximation for not very large
$l$ and for excitations with radial quantum numbers $n_r \gg l$.

For the case $l = 0$ eqs.(42)-(44) give
\be
   \mu_i (\tau) = \sqrt{\vec{p}~^2 + m^2_i} ~, ~~~~~~ i = 1, 2 ~,
\ee
\be
    \nu (\tau, \beta) = \sigma \mid \vec{r} \mid
\ee
so that the Hamiltonian (40) finally has the form
\be
   H (\vec{p}, \vec{r}~) = \sqrt{\vec{p}^{~2} + m^2_1} +
   \sqrt{\vec{p}^{~2} + m^2_2} + \sigma \mid \vec{r} \mid ~,
\ee
where $\vec{p}^{~2} = \vec{p}^{~2}_r = \frac{(\vec{p}\vec{r}~)^2}
{\vec{r}^{~2}}$.

We note again, that in the case of $l = 0$
the string doesn't contribute into the kinetic terms and
is responsible for the inert potential term $\sigma \mid
\vec{r} \mid$.

Expression (54) (valid strictly speaking only for $l = 0$)
is widely used in the context of the so-called "relativistic quark
models" [11] for arbitrary $l$. The approximate version of this
Hamiltonian was derived in ref.[7] .

Now we consider the case of small values of $l$ or $l \ll
n_r$. To develop a perturbation expansion for the spectrum
it appears convenient  to recover
in expression (40) the dependence on the total momentum
$\vec{p}^{~2} = (\vec{p}^{~2}_r + L^2/ \vec{r}^{~2})$ with the result for $H$
$$ H (p, r, \nu, \mu_i) = \frac{1}{2} \left[  \mu_1 + \mu_2
   + \frac{\vec{p}^{~2} + m^2_1}{\mu_1} + \frac{\vec{p}^{~2} + m^2_2}
   {\mu_2} + \int\limits^1_0 \frac{\sigma^2 d \beta}{\nu}
   \vec{r}^{~2} +  \right.  $$
\be
   \left. + \int\limits^1_0 d \beta \nu + \left\{ \vec{L}^2 /
   \vec{r}^{~2} \left(  \frac{1}{\mu_1 (1-\zeta)^2 + \mu_2 \zeta^2
   + \int\limits^1_0 d \beta (\beta-\zeta)^2 \nu }  - \frac{1}
   {\tilde{\mu}}  \right)   \right\}  \right]
\ee
where $\frac{1}{\tilde{\mu}} = \frac{1}{\mu_1} + \frac{1}{\mu_2}$~.

As well as in [1] we will demonstrate now that for not very large values of $l$
(or for $l \ll n_r$) the term in curly brakets corresponding to the string
contribution  to the (orbital) kinetic part of the expression (40)

\be
   H^{(1)} = \hat{\vec{L}}^2 /\vec{r}^{~2} \left(
   \frac{1}{\mu_1 (1-\zeta)^2 + \mu_2 \zeta^2
   + \int\limits^1_0 d \beta (\beta-\zeta)^2 \nu }  - \frac{1}{\tilde{\mu}}
   \right)
\ee
can be treated as a perturbation to $H^{(0)} = H - H^{(1)}$.

In zero approximation one again obtains
\be
   \mu^{(0)}_1 = \sqrt{\vec{p}^{~2} + m^2_1} ~, ~~~~~
   \mu^{(0)}_2 = \sqrt{\vec{p}^{~2} + m^2_2} ~,
\ee
$$   \nu^{(0)} = \sigma \mid \vec{r} \mid .  $$
and
\be
   H^{(0)} = \sqrt{\vec{p}^{~2} + m^2_1} +  \sqrt{\vec{p}^{~2} + m^2_2}
   + \sigma \mid \vec{r} \mid
\ee
where $ \vec{p}^{~2} = (\vec{p}^{~2}_r + \vec{L}^{~2} / r^2) $.

This potential like regime (small $l$) corresponds
(for $m^2_i \lsim \sigma $) to
\be
   < ~\mu_i~ > ~ \sim ~ ~\frac{1}{2}~ < ~\nu~ >
\ee
with nearly pure inert contribution from the
string. It is important  that even for $m_i = 0 $  the
quark-string interaction generates dynamical quark masses
$\mu_i \sim \sqrt{\sigma}$.

The first order correction $\varepsilon^{(1)}$ is determined by the average
of $H^{(1)}$ over wave functions of Hamiltonian (58)  and can be represented
in the form
\be
\varepsilon^{(1)} =
   < \frac{1}{2}  \frac{\sigma^2 l (l + 1)} {\nu^{(0)^2}} \left(
   \frac{1}{\tilde{\mu}'^{(0)} - \frac{1}{6} \nu^{0}} -
    \frac{1}{\tilde{\mu}^{(0)}} \right)  >
\ee
where we have introduced
\be
\tilde{\mu}^{(0)} = \frac{\mu^{(0)}_1
   \mu^{(0)}_2} {\mu^{(0)}_1 + \mu^{(0)}_2} ~; ~~~~~~~~~~~~
   \tilde{\mu}'^{(0)} = \frac{\mu'^{(0)}_1 \mu'^{(0)}_2}
   {\mu'^{(0)}_1 + \mu'^{(0)}_2 } ~,
\ee
with $\mu'^{(0)}_i = \mu^{(0)}_i + \frac{1}{2} \nu^{(0)}$.

To separate problems we first consider two limiting cases:
heavy-light meson ($h. l$) with
\be
   \mu_2 = m_2 \to \infty ~, ~~~~~~~~~~ m_1 = 0
\ee
and light-light meson ($l.l$) with
\be
    m_1 ~=~ m_2 ~=~ 0 ~.
\ee
where a unifyed description can be developed.
Effects connected to the presence of finite masses will be discussed
in subsection D.

In these two limiting cases zero order Hamiltonian (58) takes a simple form
[11]
\be
   H^{(0)} - m_2 = K \sqrt{\vec{p}^{~2}} + \sigma \mid \vec{r} \mid
\ee
where $K = 2$ for $l.l$ mesons and $K = 1$ for $h.l$ ones
(with the subtraction of the heavy mass for $h.l$ systems).
The corresponding spectra can be expressed in the form [11]
\be
   \varepsilon^{(0)}_{nl} = \sqrt{K} E^{(0)}_{nl} ~ ,
\ee
where
\be
   \left( E^{(0)}_{nl} \right)^2 = \pi \sigma
   \left(  2n_r + \frac{\lambda (n_r)}{\pi} l + \frac{3}{2} +
   \delta (n_r, l)   \right) ~ ,
\ee
with  $\lambda (n_r) \cong 4$ and $\delta (n_r, l)$ is a small correction.
 Neglecting  def\/lection (56) one would get for the slope
\be
   \frac{d l}{d M^2} ~ = ~ 1/4 K \sigma
\ee
 and the results of this approximation (for the leading trajectory)
are shown as crosses in Fig.1.
As we shall see def\/lection (56) from the "relativistic" Hamiltonian
(58) gives rise to a change $\sim 25\%$  in slopes of Regge
trajectories for (heavy-)light mesons, so that they are very close
to the asymptotical string slopes  $1/ K\pi\sigma$ from the very begining
of the trajectories.

To calculate  matrix elements of $H^{(1)} = H - H^{(0)}$ over the wave
functions of Hamiltonian (64) we will neglect  as well as in [1,2]
the dispersion of
$\mid \vec{p} \mid$ and $\mid \vec{r} \mid$, i.e. will perform the substitution

\be
< \mid \vec{p} \mid ^{m} \cdot \mid \vec{r} \mid^{n} > ~ \longrightarrow
{}~ < \mid \vec{p} \mid >^{m} \cdot < \mid \vec{r} \mid >^{n}
\ee

In Appendix B  it is shown that
within the accuracy $\sim 10\%$ the value of $\varepsilon^{(1)}$
can be calculated in the form
\be
   \varepsilon^{(1)}_{nl} ~ = ~ \sqrt{K} E^{(1)}_{nl} ~ ,
\ee
where
\be
   E^{(1)}_{nl} ~ = ~- \frac{\sigma^2 l (l + 1)}{E^{(0)^3}_{nl}} ~.
\ee

Therefore in this approximation one gets for not very large $l$
\be
   \left( \varepsilon_{nl} - m_2  \right)^2  =
   K \left( E^{(0)}_{nl} + E^{(1)} \right)^2
\ee
(with  disregard of the term $\sim (\varepsilon^{(1)})^2,$ since it exceeds the
accuracy of the first order calculations)
and the corresponding predictions for $(\varepsilon_{nl} - m_2)^2
/ K \pi \sigma$ are shown by open circles in Fig.1. For the leading
trajectory and not very large $l ~(l \lsim 5)$
\be
     \varepsilon^{(1)} / \varepsilon^{(0)}  ~\lsim~  0.05 ~,
\ee
with the decreasing effect for the dauther trajectories. But we stress
again that as for the change in the slope this correction leads to a
deflection of the order  $\sim 20-25\%$ as compared with pure
potential result (67).

To observe the transition into another regime we consider the
dependence of solution (44)  for $\nu(\tau, \beta)$  on $\beta$.
Taking for simplicity the case $m_1 = m_2 \le \sqrt \sigma$ (and therefore
 $\mu_1 = \mu_2, \zeta = 1/2$) one can estimate the term $\sim \vec{L}^2$
in eq. (44) as one giving
a correction to $\nu^{(0)}$ of eqs.(57)
(in the same way as we have done for eq. (60) )~ with the result
\be
 < \frac{l (l+1)}{a^2_3} > (\beta -
   1/2)^2 \approx \frac{36 l (l+1)}{(\pi (2n_r + \frac{4}{\pi} l +
   \frac{3}{2}))^2} (\beta - 1/2)^2 ~.
\ee
In the limit $l \gg n_r$ comparing the unity in right hand side of eq.(44)
with the asymptotical value of expression (73)
$$   2 (\beta - 1/2)^2$$
we encounter a large deflection (due to increase of the string contribution
to the orbital momentum) from zero order equation leading to expr.(57).
This  transition is governed obviously by the parameter
\be
\frac{l}{2n_r + 3/2}
\ee
To take it into account we are to start for large orbital momenta  $l >> n_r$
with another zero
approximation for Hamiltonian (40).
\vspace{5mm}


\centerline{\bf C. ~Relativistic string regime for large orbital momenta}
\vspace{3mm}

Let us consider the case of
\be
   l \gg n_r
\ee
where as we will prove light quarks in the meson carry only a small part of the
total energy (orbital momentum)
\be
<~~ \nu ~~> ~~\gg~~ <~~ \mu_i ~~>
\ee
where $i = 1, 2 $  for l.l. meson (with
$m_1, m_2 \lsim \sqrt{\sigma}$)~ and $i = 1 ~
(m_1 \sim \sqrt{\sigma}, \mu_2 = m_2 \to \infty)$ for h.l. meson. For equal
masses case $m_1 = m_2$ it was shown in ref.[1].

For $l \gg n_r$ one can exploit the quasiclassical condition
\be
   < (r - < r > )^2 > ~ \ll ~ < r >^2
\ee
and expand [1] the $r^2$-depending part of Hamiltonian (40) around
the extremum
\be
   r^2_l = \left(    \frac{l(l+1)(\sigma^2 \int d \beta/\nu)^{-1}}
   {(\mu_1 (1-\zeta)^2 + \mu_2 \zeta^2 + \int\limits^1_0
   d \beta (\beta - \zeta)^2 \nu) }  \right)^{1/2}
\ee
so that in gaussian approximation
\be
    H = \frac{1}{2} \left[ \frac{p^2_r + m^2_1}{\mu_1} +
    \frac{p^2_r + m^2_2}{\mu_2} + \mu_1 + \mu_2 + \int \nu d \beta ~ +
    \right.
\ee
$$  \left.  + ~ 2 \left( \frac{l(l+1)\int \frac{\sigma^2 d \beta}{\nu}}
    {\mu_1 (1-\zeta)^2 + \mu_2 \zeta^2 + \int(\beta-\zeta)^2 \nu d \beta}
     \right)^{1/2} + 4 \int \frac{\sigma^2 d \beta}{\nu}
    (r - r_l)^2     \right] ~. $$

As in the case of small $l$ we first concentrate on the two limiting
cases (62), (63).
For $m_1 = m_2$ the symmetry between quarks allows one [1]
to restrict the class of auxiliary fields by
\be
   \mu_1 = \mu_2 ~, ~~~~~~~~~ \int (\beta - 1/2) \nu d \beta = 0
\ee
and consequently
\be
   \zeta ~ = ~ 1/2 ~.
\ee
For~  $\mu_2 \to m_2 \to \infty$ one obviously gets in the leading order
\be
   \zeta = 0 ~, ~~~~~~~~~\mu_2 = m_2 ~ .
\ee

Let perform in the limit (75)
the expansion of  Hamiltonian (79) in powers of $\mu/\nu$.
In zero approximation one gets (neglecting radial dynamics,
$m_i$ and $\mu_i$ of light quarks)
\be
   \left( H^{(0)} - m_2  \right) = \frac{1}{2} \int \nu d \beta +
   \left( \frac{\sigma^2 l (l+1) \int \frac{d \beta}{\nu}}
    {\int (\beta - (K-1)/2)^2 \nu d \beta}   \right)^{1/2} ~ ,
\ee
where $K = 2$ for l.l. case and $K = 1$ for h.l. one.
We emphasize, that in this approximation one recovers the pure
straight-line string Hamiltonian [12, 1] without radial excitations,
which will appear only as a correction to expression (83).

The extremal value of $\nu^{(0)}(\beta)$ one can find in the same
way as it has been done for l.l. case in [12, 1].
\be
   \nu^{(0)} = \left( \frac{4K \sigma \sqrt{l(l+1)}}{\pi} \right)^{1/2}
   \frac{1}{\sqrt{1 - \left( K \left( \beta - \frac{K - 1}{2}
   \right) \right)^2 } } ~.
\ee
This energy
distribution along the string leads to the following total energy in zero
approximation
\be
   \left( \varepsilon^{(0)} \right)^2 = K \pi \sigma \sqrt{l(l+1)} ~ .
\ee
Expression (84) for the energy density can be easily interpreted
as a rotation of homogeneous distribution (57) of potential case
if one recognises that  factor
$$  \frac{1}{\sqrt{1 - \left( K \left( \beta - \frac{K-1}{2}
    \right)^2 \right) }}  ~ =  ~ \frac{1}{\sqrt{1-v^2(\beta) } }  $$
represents the standard Lorentz-factor, with $v(\beta) = K (\beta -
\frac{K-1}{2} ) $
playing the role of the velocity of the corresponding elementary piece of the
string.

For l.l. system
\be
   - v (0)~ = ~v (1)~ = ~1
\ee
and for h.l. system  one has one half of the string
$$ v(0) = 0~~, ~~~v(1) = 1  ~ .$$

Let us find the leading correction $H^{(1)}$ to zero approximation
 (83) and recover our starting condition $< \mu_i >
\ll < \nu > $  (for $m^2_i \lsim \sqrt{\sigma}$).
Again we first discuss two limiting cases (62), (63) for
$$     m_1 ~ = ~ m_2 ~ = ~ 0 ~,$$
or
$$     m_1 ~ = 0 ~, ~~~~ m_2 \to \infty ~.$$

In Appendix C it is shown, that expanding eq.(79)
in $\mu/\nu$ up to the first nonvanishing terms one obtains
(after integration over $\mu$) the following expression for $H^{(1)}$
\be
   H^{(1)}_r = \sqrt{K} \left( \frac{3^{4/5}}{4} (\pi \sigma)^{1/2}
   l^{-3/10} \right) \left( \left( p^2_x +c \Sigma_i m^2_i \right)^{2/3}
   + x^2 \right) ~ ,
\ee
with
$$   \mu^{(0)} = \left( \frac{\varepsilon^{(0)}}{3 K}
    \left( p^2_r + \sum_i m^2_i \right) \right)^{1/3} ~ ,$$
where  the sum is over $i = 1,2$ for l.l. and $i = 1$ for h.l. mesons,
$c = \frac{3^{12/5}~l^{1/5}} {\pi\sigma}$  and
we have introduced a dimensionless variables
\be
\frac{(r -
r_l)}{x} = \frac{p_x}{p_r} = \sqrt{K} 3^{2/5} (\pi \sigma)^{-1/2}
   l^{1/10}~,  ~~~ p^2_x = - \partial^2_x
\ee
and as before $K = 2, 1 $  for
l.l. and h.l. systems correspondingly.  For the case of $K = 2$ this
expression was obtained in ref.[1].

Within the accuracy $\sim 5 - 10 \% $ the spectrum of $H^{(1)}_r$
can be calculated (see Appendix C for details) in the form
\be
   \varepsilon^{(1)}_{n_r, l} = \sqrt{K \pi \sigma} \left( \frac{3}{4}
    \right)^{1/5}  \frac{5}{4} \left( n_r + 1/2 \right)^{4/5} l^{-3/10}
\ee
and in accordance with our initial assumption (76) one indeed obtains
\be
   \frac{\mu}{\nu} \sim \left( \frac{n_r + 1/2}{l} \right)^{2/5}  \ll 1
\ee
in the limit (75) we are dealing with.

The resulting spectrum for $M^2/K\pi\sigma$, where
\be
   M = \varepsilon^{(0)} + \varepsilon^{(1)}
\ee
is shown by open squares in Fig.1 and doesn't depend on $K$ (again we are to
disregard the term $\sim (\varepsilon^{(1)})^2$ since it exceeds the
accuracy of the calculations). As we argue in Appendix C the exact value of
$(\varepsilon^{(1)})_{exact} \approx 0.9 \varepsilon^{(1)}$ and we take it into
account in this figure.
It follows from Figure 1
that low $l$ and high $l$ approximations join very smoothly forming
(within the accuracy $\sim 5\%$) Regge trajectories with the slope
\be
   \frac{d M^2}{d l}  ~ = ~ K \pi \sigma
\ee
and intersepts, determined from the Hamiltonian (64)
\be
      M^2 (l = 0) = K \pi \sigma \left( 2n_r + \frac{3}{2} + \delta \right) ~.
\ee
The values for $M^2$ are very close to results found numerically
in ref.[5] (black dots in Fig.1). A small deflection between them is
out of the accuracy of the first order calculations we have done. A short
discussion of similarities and differences between our approach and
that of ref.[5] is postponed till the Conclusion.

It is important to stress, that  "minimal" QCD string combines together
 properties of both the string models (string like slope (92) )
and ones of the potential models (existence of  radial excitations,
which lead to the intersept (93) ). We have found both in the l.l. and in h.l.
systems two dynamical regimes distinguished by the contribution
of the string into the (orbital) kinetic part of the action: in the potential
regime for small $l$ or
\be
   \frac{l}{2n_r + 1}  ~ \ll ~1
\ee
the string constitutes inert linear potential with $\nu(\tau, \beta)$
close to the homogeneous
distribution (57); in the string regime for
\be
   \frac{l}{2n_r + 1}  ~ \gg ~1
\ee
the main part of the energy (orbital momentum) is carryed by the
string with energy density $\nu (\tau, \beta)$  given in the leading
order by expression (84).
\vspace{10mm}

\centerline{\bf D. ~Mass effects in the minimal QCD}
\centerline{\bf string with quarks }
\vspace{3mm}

To analyse the dynamics for  arbitrary quark masses case we first consider
l.l.
mesons with
\be
  m_1 \lsim \sqrt{\sigma} ~, ~~~~~~~~ m_2 \lsim \sqrt{\sigma}
\ee
and h.l. systems with
\be
    m_1 \lsim \sqrt{\sigma} ~, ~~~~~~~~ m_2 \gg \sqrt{\sigma} ~ .
\ee

Our main aim here will be to estimate upper bound on light masses,
which don't change considerably $K\pi\sigma$ slope (92) of  Regge
trajectories for massless light quark(s), discussed in the
previous section.
 We also calculate in this case  leading mass corrections to the
spectrum.

As in the previous section we  concentrate on two limiting regimes (94), (95).

In the potential regime (94) zero approximation as well as in the previous
section is described by Hamiltonian (the subtraction of the heavy mass in
h.l. system is implyed)
\be
    H^{(0)} = \sqrt{\vec p^{~2} + m^2_1} +
    \sqrt {\vec{p}^2 + m^2_2} + \sigma \mid \vec{r} \mid
\ee
with the spectrum [11]
in the case of $m \ll < \mid \vec{p} \mid > \sim E^{(0)}$
\be
   \tilde \varepsilon^{(0)} \approx  \varepsilon^{(0)} \left( 1 +
   \frac{\sum_{i} m^2_i}{E^{(0)2}} \right) ~,
   ~~~~~~\varepsilon^{(0)} = \sqrt{K}E^{(0)} ~,
\ee
where $E^{(0)}$ is given by (66) and the sum is over $i = 1.2$ for l.l. or
$i = 1$ for h.l. mesons.

We calculate the correction $\varepsilon^{(1)} = < H - H^{(0)} >$ in the
same way as for the massless case (70) of the previous section
(see Appendix B for details).

Making use of the virial theorem and neglecting the dispersion of
$\mu_i~, \nu$ one gets in the leading order of $\left( \frac{\sum_i
m^2_i}{E^{(0)2}} \right)$

$$ < \mu^{(0)} > \approx  \frac{\varepsilon^{(0)}}{2K} \left( 1 + 3
   \left( \frac{\sum_i m^2_i} {E^{(0)2}} \right) \right)  ~ ,$$

\be
   < \nu^{0} > \approx
   \frac{\varepsilon^{(0)}}{2} \left( 1 - \left( \frac{\sum_i m^2_i}
   {E^{(0)2}} \right) \right) ~ .
\ee

The substitution of (100) in (60) gives for the correction $\varepsilon^{(1)}$
in massive case
\be
\left(  \frac{\varepsilon^{(1)}}{\varepsilon^{(0)}} \right) = -\frac{l(l+1)
\sigma^2} {E^{(0)4}} \left( 1 - 4 \frac{\sum_i m^2_i}{E^{(0)2}} \right) ~ .
\ee

Therefore we have (up to the term $\sim m^2
\frac{\varepsilon^{(1)}}{\varepsilon^{(0)}}$ ) for the
mass squared $M^2 = (\tilde{\varepsilon}^{(0)} + \varepsilon^{(1)})^2$
\be
M^2 = K \left[ E^{(0)2}  \left( 1 - \frac{l(l+1)\sigma^{2}}{E^{(0)4}}
\right)^2 + 2 \sum_i m^2_i (1 + 3 \frac{l(l+1)\sigma^2}{E^{(0)4}} ) \right] ~ .
\ee

Let us estimate quark mass corrections to the slope of the trajectory. The
first
term of eq. (102)
gives in accordance with  expression (92)
$$  \frac{dM^2}{dl} \approx K \pi \sigma ~ . $$
For the leading Regge trajectory
the deflection of the slope from the massless case can be expressed
in the form
\be
\frac{\Delta \frac{dM^2}{dl}} {K \pi \sigma} \approx \frac{8}{3 \pi^3}
f ( l) \frac{\sum_i m^2_i}{\sigma}
\ee
where $f ( l) = \frac{1}{(1  + \frac{8l}{3\pi})^2}$
with $f(1) \approx 4$.

Hence the slope of  leading Regge trajectory is not essentialy changed when
\be
\sum^K_{i=1}m^2_i \ll \frac{3\pi^3}{2} \sigma \sim 40\sigma
\ee
where $K = 2$ for l.l. and $K = 1$ for h.l. system. For the daughter
trajectories the restriction for the masses is even weaker.

We emphasize that restriction (104) is much more moderate as compared with
that for the deflection of intersept (93) (the second term in the r.h.s.
of eq.(99)) to be small, which amounts to
\be
   \sum^K_{i = 1} m^2_i ~ \ll ~ \frac{3\pi}{2} \sigma ~.
\ee

To complete analysis let us consider the case $l \gg n_r$
when  radial quark dynamics can be considered as a perturbation [1].

We start with  expression (87) for the Hamiltonian $H^{(1)}_r$ which
determines the corrections to zero (pure string) approximation (83)-(85). As
it is shown in Appendix C the string asymptotics
(where $< \nu > \gg < \mu_{light} > $) is valid for
\be
\frac{l}{n_r + \frac{\sum_{i=1}^K m^2_i}{\pi\sigma} + \frac{1}{2}}
\gg 1 ~ .
\ee
Keeping this condition satisfied we first consider the case  of nonrelativistic
radial dynamics when $p^2_r \ll
\sum_i m^2_i$~, which is achieved as we shall see if
\be
\frac{l}{(n_r + \frac{1}{2})^6} ~ \gg ~ \left( \frac{\pi\sigma}{\sum_i m^2_i}
\right) ^5 ~.
\ee

In this limit one obtains (see Appendix C for details) the spectrum of
$H_r^{(1)}$ in the form
\be
\varepsilon^{(1)}_{n_r, l} = \sqrt{K} \left( {\frac{3}{4}}^{4/3}
\left( \sum_i m^2_i \right)^{2/3} (\pi\sigma l)^{-1/6} + \left(
\frac{3}{8}\right)^{1/6}
\frac{(\pi\sigma)^{2/3}}{(\sum_i m^2_i)^{1/6}} l^{-1/3} \left(n_r +
\frac{1}{2}\right) \right) ~ .
\ee

It is easy to make sure, that condition (106) is equivalent to
$\varepsilon^{(1)}_{n_r, l} / \varepsilon^{(0)}_l \ll 1$.
Making comparison between $p^2_r$ and $\Sigma_im^2_i$ one recovers that
in the regime (106), (107)  radial dynamics is indeed "nonrelativistic"
\be
   < p^2_r > \sim (\sum_i m^2_i (\pi\sigma)^5 )^{1/6} l^{-1/6}
   (n_r + \frac{1}{2}) \ll   \sum_i m^2_i ~ .
\ee

We note that condition (109) follows also from the requirement,
that the first term of eq.(108) is much more than the second one.

To calculate the spectrum in the opposite asymptotics of string regime (106)
\be
   < p^2_r > ~ \gg ~ m^2
\ee
and merge smoothly the massless case (89) one is to consider (see Appendix
C)  the following domain of $l, ~n_r$
\be
  \frac{l}{(n_r + \frac{1}{2})^6} ~ \ll ~ \left( \frac {\pi\sigma}{\sum_i
m^2_i}
   \right)^5 ~ ,
\ee
where the spectrum of $H^{(1)}_r$ has the form
\be
    \varepsilon^{(1)}_{n_r, ~l} =
    \sqrt{K} \left( \left( \frac{3}{4} \right)^{1/5}~~
    \frac{5}{4} (\pi \sigma)^{1/2}  l^{-3/10} (n_r + 1/2)^{4/5} ~ + \right.
\ee
$$  \left. + ~
    \left( \frac{3}{4}\right)^{2/5} \sum\limits_i m^2_i
    (\pi\sigma)^{-1/2}~l^{-1/10}  (n_r + 1/2)^{-2/5}  \right) ~ . $$

Restriction (110), (111) comes from the condition, that
the first term of eq. (112) (which is the massless correction
(89) of the previous section) is much more, than the second one.
Due to condition (106) we again have the restriction
$\varepsilon^{(1)}_{n_{r},l} / \varepsilon^{(0)}_l \ll 1$
to be satisfyed.

To conclude this section
let us discuss briefly the transition from nonrelativistic dynamics of heavy
quark(s) (45) in the heavy-heavy or heavy-light mesons at small $n_r~, l$ to
the
large $n_r~, l$ regime, when both quarks become relativistic. We concentrate
on the connection between this transition and the transition from the
potential regime of subsection A to the string one.
As we easily conclude from eq. (66) the
relativization, ~$\vec{p}^2 \sim m^2$,~ appears for
\be
   \left( n_r + \frac{2}{\pi} ~l\right) ~ \sim ~ \frac{2}{\pi}
    \frac{(m^2_1 + m^2_2)}{\sigma} ~ .
\ee

If  excitation of the system corresponds to $n_r \gg l$
, then one achieves the potential relativistic dynamics, described by the
Hamiltonian (58) with the small string corrections (60). In the opposite case
of
orbital relativization $l \gsim n_r$ as it is clear from the discussion of
eq. (73) there is a considerable diflection from pure potential
regime (58). Therefore in this case both quarks start to become relativistic in
the
regime intermediate between potential one
(58) and string one (83)
(valid under the condition  (106) )
where the dependence of $\nu$ on $\beta$ is considerable but still
dif\/fers from the asymptotical one (84).
\vspace{7mm}


{\bf 6 ~CONCLUSIONS}
\vspace{5mm}

Let us summarize the results. We obtain the generalization of
Hamiltonian [1] for the spinless quark and antiquark in the confining QCD
vacuum
 for the case of arbitrary quark masses. Starting from the QCD Lagrangian
we make use the minimal area law asymptotics for the averaged Wilson loop
which leads to the appearence of the minimal QCD string connecting quarks.
The string contributes (for $l>0$) in the kinetic part of the effective
action and therefore in order to introduce the center of masses and relative
coordinates we are to use explicitly the condition that they are decoupled
from each other. Introducing the auxiliary fields we represent the action in
the quadratic form with respect to the quark coordinates and
diagonalization of the kinetic part of the action leads to the proper
definition of the total
momentum $\vec P$. Choosing the rest frame of the meson $\vec P=0$ we
arrive (neglecting pathes of quarks with backtracking in time) at the
effective Hamiltonian. Additionally to  quark coordinates it contains
the auxiliary fields playing the roles of the quark dynamical masses
$\mu_i(\tau)$ and the string energy density $\nu(\tau,\beta)$. Integration
over them amounts to the substitution of their extremal values after which
one is to construct the operator of Hamiltonian acting on the wave functions.

   The interaction between  quarks and the string gives rise to
   appearence of  two different dynamical regimes (as well as for the equal
masses
case [1]), which are distinguished by the string contribution into the
kinetic part of the Hamiltonian. For the low orbital momenta ($l \ll n_r$)
dynamics
is described in the leading order by the  relativistic linear potential
Hamiltonian with almost inert contribution from the string
constituting  potential $\sigma \mid \vec{r} \mid$. In the opposite limit
of $l \gg n_r$ the system behaves as the rotating  string which carries the
main part of the orbital momentum and  energy.  The transition between
these regimes is smooth and  relativization of heavy quark(s) due to the
increase of $l$ is also considered.

We develope the unifyed description of
the heavy-light and light-light mesons and prove that in these limiting cases
Regge trajectories are nearly straight line with the slope close to
$1/K \pi\sigma$ $(K = 1,2$ for h.l. and l.l mesons respectively). The upper
bound on the light mass(es) which don't change considerably the slope is
obtained and the leading mass corrections are calculated for both regimes.
It appears that the slope is much less sensitive to increase of quark masses
as compared with the intersepts of the trajectories.

We note that our results for the spectrum are very close to that obtained by
the numerical quantization of the same action [5] postulated without
 derivation from QCD Lagrangian.  We stress that as compared with [5] our
approach gives the qualitative picture of underlieing regimes of
quark-string interaction. Also it enables one to avoid in the leading order
the complications of Weil ordering [9] in Hamiltonian (40). In zero
approximation Hamiltonian both for small $l ~(l \ll n_r)$ (58) and for $l
\gg n_r$ (83) is the sum of the terms depending either only on the
momentums or space coordinates which doesn't require any Weil ordering. The
need for this ordering appears only for the correction terms (56), (87) and
calculation of them can be performed with the neglect of the dispersion
(Weil ordering).  Such approximate procedure gives the value for the leading
corrections within more than 10\% of accuracy.  In [5] the numerical
procedure makes it difficult to keep track of the ordering and relies on the
smallness of the dispersion (see forthcoming paper for a more detailed
discussion).

We also consider the modification of the heavy quarkonia Hamiltonian arising
from the string contribution (at distances  $\gsim T_g$)  to the orbital
momentum. We estimate the energy correction due to this effect and find
conditions under which it is comparable with the first relativistic correction.

The authors are grateful for usefull discussions to Yu.S.Kalashnikova,
   A.B.Kaidalov, Yu.A.Simonov, J.A.Tjon and M.G.Olsson.
   They would like also to thank N.A.Aksenova for typing this manuscript.
   This work is supported by Russian Fundamental Research Foundation,
   grant N 93-02-14937.


\newpage
\setcounter{equation}{0}
\renewcommand{\theequation}{A.\arabic{equation}}

{\bf Appendix ~A}

In this Appendix we perform the gaussian integration (32) over $\eta (\tau,
\beta)$ to go over from effective action (31)
\begin{equation}
\tilde{A}  = \frac{1}{2} \int\limits^{T}_{0}~d\tau \left[
\frac{m^2_1}{\mu_1} + \frac{m^2_2}{\mu_2} + \frac{1}{a_1}
\{ a_3 a_1 \dot{\vec{r}}^2 - 2c_2a_1 (\vec r\dot{\vec{r}}) +
(a_4 a_1 - c_1^2)\vec r^2 + a_1^2
\} \right]
\end{equation}
to eq.(34) expressed in terms of
physical quantities: dynamical masses $\mu_1, \mu_2$  energy density
along the string $\nu (\tau, \beta)$ and relative coordinates.

In what follows we make use of the fact that  gaussian integration
effectively amounts to
substitution of the extremum value of $\eta^{ext}$ into eq.(A.1). On the  basis
of the
definitions (21) of $a_i$ and $c_i$  we obtain from (A.1) the following
extremum condition for $\eta$.
\begin{equation}
2 \int\limits^{1}_{0} d\beta\nu \left[ r^2(a_1 \eta - c_1)  -
(r\dot{r}) a_1 (\beta - \zeta) \right] \delta\eta = 0
\end{equation}
that results in
\begin{equation}
\eta^{ext}  = \frac{c_1}{a_1} +  \frac{(r \dot{r})}{r^2} (\beta-\zeta)
\end{equation}
with $c_1 = \int\limits^{1}_{0} d\beta \eta^{ext} \nu$.

Resolution of (A.3) finally gives
\begin{equation}
  \eta^{ext}  = \frac{(r\dot{r})}{r^2}
  \left( \beta - \frac{\mu_1}{\mu_1 + \mu_2} \right) ~ .
\end{equation}
Expression (A.4) leads  to  the following $c_i^{ext}$
\begin{equation}
c^{ext}_1  = \frac{(r\dot{r})}{r^2} (\zeta - \frac{\mu_1}{\mu_1 + \mu_2})~a_1
\end{equation}

$$c^{ext}_2  = \frac{(r\dot{r})}{r^2} (a_3 - \mu_1 + \frac{\mu_1^2}{\mu_1 +
\mu_2})  ~ ,$$
where we make use of
$$
\int\limits^1_0 d\beta\nu  = a_1 -  (\mu_1 + \mu_2) ~ ,
$$
\be
\int\limits^1_0 d\beta\nu\beta = \zeta a_1 - \mu_1 ~ ,
\ee
$$
\int\limits^1_0 d\beta\nu\beta^2 = a_3 - \mu_1 + \zeta^2 a_1 ~ .
$$
To calculate (A.1) with $\eta = \eta^{ext}$ we note that for
quadratic in $\eta$ form one has the following relation at the extremum (A.2)
\begin{equation}
     (\tilde{a}^{ext}_4 a_1 - (c^{ext}_{1})^2 ) ~r^2 = (r\dot{r})
     c_2^{ext} a_1 ~ ,
\end{equation}
where we have  separated $\eta$- dependent part of $a_4$ introducing
$\tilde{a}_4 = \int\limits^1_0 \eta^2 \nu d \beta$.

With the help of (A.7) one concludes that $\eta$-dependent part of the
action (A.1) amounts to the following
contribution into $\tilde{A}$
\begin{equation}
   - (r \dot{r}) c_2^{ext}~a_1 ~ ,
\end{equation}
that results finally in the  effective action (34)  for  the rest frame
\begin{eqnarray}
A' (\mu_1, \mu_2, \nu) = \frac{1}{2} \int\limits^{T}_{0}~d\tau \left[
\frac{m^2_1}{\mu_1} + \frac{m^2_2}{\mu_2} + \mu_1 + \mu_2 + (\mu_1
(1-\zeta)^2 + \mu_2 \zeta^2 + \int\limits^1_0 d\beta (\beta-\zeta)^2 \nu)
\times \right. \nonumber \\
\left. \times \frac{[\dot{\vec{r}} \times \vec{r}]^2}{\vec{r}^2} +
\frac{(\dot{\vec{r}} \vec{r})^2} {\vec{r}^2} \tilde{\mu} + \vec{r}^2
\int\limits^1_0 d\beta \frac{\sigma^2}{\nu} + \int\limits^1_0 d\beta \nu
\right]~~~~~~~~~~~~~~~~~~~~~
\end{eqnarray}
where $\tilde{\mu} = \frac{\mu_1 \mu_2}{\mu_1+\mu_2}$ is the reduced
dynamical mass.


\newpage

\setcounter{equation}{0}
\renewcommand{\theequation}{B.\arabic{equation}}

{\bf Appendix B}
\vspace{5mm}

In this Appendix we calculate first order correction (56) to the
relativistic Hamiltonian (58). Making use of the approximation (68),
proposed in [1], we neglect in (56)
the despersion of $~|~ \vec p ~|~$ and
$~|~ \vec r ~|~$ (and disregard also
Weil ordering [9]) and express $\varepsilon^{(1)} = < H^{(1)} > $
in terms of $ < ~|~ \vec p ~|~ > , ~< ~|~ \vec r ~|~ > $
averaged over wave functions of zero order Hamiltoian (58) and
connected to $E^{(0)}$
of eq.(66). For this purpose we first evaluate $< \mu^{(0)}_i > ,~
< \nu^{(0)} > $ (determined by eqs.(57) ) for $m^2_i \lsim
\sqrt{\sigma}$ and then substitute them to  eq.(60) for
$\varepsilon^{(1)}$. Also we will develope the procedure to estimate the
 accuracy of this approximation
 and find that  account of nonzero dispersion (Weil ordering)
 would change the result for $\varepsilon^{(1)}$ less then
for 10\%.

First we use the virial theorem for zero order Hamiltonian (58)
\be
   \left< \frac{\vec p^{~2}}{\sqrt{\vec p^{~2} + m^2_1}}
   + \frac{\vec p^{~2}}{\sqrt{\vec p^{~2} + m^2_2}}  \right> =
   \sigma  < ~|~ \vec r ~|~ >
\ee
where the average is performed over the wave functions of (58).

Neglecting the dispersion one easily obtains the system of
equations for $ < \mu^{(0)}_1 > , \\
< \mu^{(0)}_2 > $~ and ~$< \nu^{(0)} > $
\be
\left\{
\begin{array}{lll}
   < \mu^{(0)}_1 >  - \frac{m^2_1}{< \mu^{(0)}_1 >}  +
   < \mu^{(0)}_2 >  - \frac{m^2_2}{< \mu^{(0)}_2 >}  =
   < \nu^{(0)} >   \\
   \varepsilon^{(0)}_m = < \mu^{(0)}_1 > + < \mu^{(0)}_2 >
   + < \nu^{(0)} >  \\
   < \mu^{(0)} >^2 - m^2_1 = < \mu^{(0)}_2 >^2 - m^2_2        \\
\end{array}
\right.
\ee
where the first one follows directly from (B.1), the second one is just the
definition of the  eigenvalues of eq.(58) and the last equation ensures the
condition that the total momentum of the meson is equal to zero.

To obtain from eqs.(B.2) the values of $< \mu_i^{(0)} >$ and $< \nu^{(0)} >$
with the account of leading mass corrections $\sim\sum^K_{i=1} m^2_i/\sigma$
it is sufficient to consider
the reduced system of equations in two limiting cases of h.l. (62)
and l.l. (63) mesons
\be
   < \mu^{(0)} > - \frac{\sum^K_{i=1} m^2_i}{K< \mu^{(0)} >} =
     \frac{< \nu^{(0)} >}{K}
\ee
$$   \sqrt K E^{(0)} \cdot \left( 1 + \frac{\sum^K_i m^2_i}{E^{(0)2}}
      \right) = K< \mu^{(0)} > + < \nu^{(0)} > $$
where $ K = 2$ for l.l. system and $K = 1$ for h.l. one. The solution of the
system (B.3) gives
\be
\begin{array}{ll}
  < \mu^{(0)}  > ~\approx~ \frac{\varepsilon^{(0)}}{2K}
  \left( 1 + 3 \left( \frac{\Sigma_i m^2_i}{E^{(0)^2}}
  \right) \right)   \\
  < \nu^{(0)} > ~\approx~ \frac{\varepsilon^{(0)}}{2}
  \left( 1 -  \left( \frac{\Sigma_i m^2_i}{E^{(0)^2}}
  \right) \right)   \\
\end{array}
\ee
where $\varepsilon^{(0)} = \sqrt{K} E^{(0)}$, defined in (66).

Substitution of these values into the eqs.(60), (61) results in the following
value for \\ $< H^{(1)} > = \varepsilon^{(1)}$
\be
   \frac{\varepsilon^{(1)}}{\varepsilon^{(0)}} = -
   \frac{l(l+1)\sigma^2}{E^{(0)^4}} \left(1 - 4 \frac{\Sigma_im^2_i}
   {E^{(0)^2}} \right)
\ee
where the summation over $i$ runs from $1$ to $K$.

In the rest of this appendix we estimate the effect of  dispersion
for the calculation of $\varepsilon^{(1)}$ and
find it to be less then 10\% even for the ground state of the
Hamiltonian (58). For this purpose we first represent  Hamiltonian  (58)
(taking for the sake of definitness $K = 1$) in the form containing
auxiliary fields $\mu(\tau)$ and $\nu(\tau)$
\be
   H = \frac{1}{2}\left( \frac{1}{\mu} (\vec p^{~2} + m^2) + \mu \right)
   + \frac{1}{2} \left(\frac{\sigma^2}{\nu}\vec r^{~2} + \nu \right) ~.
\ee
Integration over $\mu(\tau)$ and $\nu(\tau)$  amounts effectively [1]
(as in the main text)
to the insertion of their extremal values
\be
   \mu_{ext} = \sqrt{\vec p^{~2} + m^2} ~,
   ~~~~~ \nu_{ext} = \sigma ~|~ \vec r ~|~
\ee
which enables to recover eq.(58) for h.l. system.

To evaluate the effect of dispersion it is sufficient to
restrict the integrations $D \mu D \nu$  by that over $\mu$
and $\nu$ independent of time.
Actually such approximation gives (for not very
large quantum numbers $l, n_r$)  def\/lection in the spectrum
(and therefore in  dispersion) less then 5 -- 10 \% (see [3] for the
discussion).  As a result within this accuracy we reduce  initial
Hamiltonian (58)
to one of the oscillator type.
Our strategy is to compare matrix elements $< | \vec p | ^n
| \vec r | ^m >$ with their approximate counterparts
$< \mid \vec p \mid >^n < \mid \vec r \mid >^m$ where averaging will be
performed over our oscillator version of eq.(58). For this purpose one
is to obtain [3] first the energy $E (\mu, \nu)$ as the function of $\mu$
and $\nu$. After this we find the extremal values of $\mu$ and $\nu$
from the conditions
\be
   \frac{\partial E}{\partial
   \mu} \left( \mu^{ext}, \nu^{ext} \right) ~ = ~ 0 ~, ~~~~~~ \frac{\partial
   E}{\partial \nu} \left( \mu^{ext}, \nu^{ext} \right) ~ = ~ 0 ~,
   \ee
and substitute them into the wave functions in order to calculate the
matrix elements under consideration.

Along these lines we have for the spectrum of the
Hamiltonian (B.6)
\be
   E(\mu, \nu) = \frac{m^2}{2\mu} + \frac{\mu}{2} + \frac{\nu}{2}
   + \varepsilon_n
\ee
$$ \varepsilon_n = \frac{\sigma}{\sqrt{\mu \nu}}
   \left(n + \frac{3}{2} \right) $$
so that
\be
   \mu^{ext} = \sqrt{\left(n + \frac{3}{2}\right) \sigma},
   ~~~~~ \nu^{ext} =
   \sqrt{\left( n + \frac{3}{2} \right) \sigma},
   ~~~~~ \omega^{ext} =
   \sqrt{\frac{\sigma}{\left( n + \frac{3}{2} \right) } }
\ee
and finally in zero order of $m^2$ one obtains the following spectrum
\be
   E^{(0)2} = 4 \left( n + \frac{3}{2} \right) \sigma ~,
\ee
with  $ n = 2 n_r + l $~ .
We note that expression (B.11) is reasonably close to the exact one
(66) ($K = 1$) and even gives the correct Regge slope (67).

The wave functions of   the approximate version of
Hamiltonian (B.6) in the momentum space have the standard form
\be
   a_{nlm} (p) = C p^l exp \left( - \frac{\alpha^2 p^2}{2} \right)
   Y_{lm} (\theta, \varphi) F \left(- \frac{n-l}{2}, l + \frac{3}{2},
   \alpha^2 p^2 \right) ~ ,
\ee
where $\alpha = \left( \frac{\sigma^2 \mu^{ext}}{\nu^{ext}} \right)^{1/4}$
and $C$ is the normalization constant.

To take the symplest example
let us estimate the dispersion contribution to the evaluation of the
first order
correction $\Delta \varepsilon \sim \left( \frac{m^2}{p} \right)$  in the
relativistic expansion of the energy
$\tilde{\varepsilon}_{nl}^{(0)}$ for a heavy-light meson
(neglecting the string contribution (56) to the orbital momentum).
Keeping only the first term in the expansion one starts with
\be
   \tilde{\varepsilon}^{(0)}_{nl} = < \sqrt{\vec p^{~2} + m^2} + \sigma
   ~|~ \vec r ~|~ > ~\approx~ < \sqrt{\vec p^{~2}} + \sigma ~|~\vec r ~|~
   > + \frac{1}{2} < \frac{m^2}{\sqrt{\vec p^{~2}}} > \equiv
   \varepsilon^{(0)}_{nl} + \Delta \varepsilon
\ee
where $\varepsilon^{(0)}_{nl}$ is given by eq.(64) with $K = 1$.
One is to compare now two quantities $1/< \sqrt{\vec p^{~2}} >$ and
$< 1/ \sqrt{\vec p^{~2}} >$.

For this purpose we consider $n_r = 0$, since in this case
the discrepancy (for a given $l$) maximal.
For such quantum numbers wave function (B.12)
is reduced with the help of Cummer formula to the simple form
\be
   a_{nlm} (p) = C \cdot p^l exp \left( -\frac{\alpha^2 p^2}{2} \right)
   Y_{lm} (\theta, \varphi)
\ee
with $ C^2 = \frac{2 \alpha^{2l+3/2}} {\Gamma \left( l + \frac{3}{2} \right) }
$.
The dispersion contribution $\Delta \varepsilon^D $ is defined
in the following way
\be
   \Delta \varepsilon = \frac{1}{2} m^2 < \frac{1}{\mu^{(0)}} >
    = \frac{m^2}{E^{(0)}} + \frac{1}{2} m^2 \left( < \frac{1}{\mu^{(0)} }
    > - \frac{1}{< \mu^{(0)} > } \right) \equiv \Delta \varepsilon
    ^{D=0} + \Delta \varepsilon^D
\ee
with $ E^{(0)} = 2 < \mu^{(0)} > = 2 < \sqrt{\vec p^{~2}} >$  given by (B.11).
The difference $\Delta\varepsilon^D$ between corresponding matrix elements
can be approximately evaluated with the help of
wave functions (B.14) with the result
\be
\begin{array}{ll}
  \frac{1}{< \sqrt{\vec p^{~2}}  >}  ~\approx~  \frac{1}{\sqrt{< \mu^{(0)} >
  \cdot < \omega^{(0)}> }} \cdot \frac{\Gamma \left( l + \frac{3}{2}
  \right) }{\Gamma ( l + 2 ) }    \\
\left< \frac{1}{\sqrt{\vec p^{~2}} } \right> ~\approx~ \frac{1}{\sqrt{<
\mu^{(0)} >
  \cdot < \omega^{(0)} > }} \cdot \frac{\Gamma ( l + 1) }
   {\Gamma \left( l + \frac{3}{2} \right) } \\
\end{array}
\ee
where $\mu^{(0)}, \omega^{(0)}$ are defined in (B.10).

Making use of Vallis formula for gamma functions
(which is valid strictly speaking
for $l \gg 1$, but still gives a required accuracy for $l \gsim 1 $ )
we obtain for the dispersion contribution
\be
   \Delta \varepsilon^D ~ = ~ \frac{m^2}{\sqrt{\sigma}}
   \cdot \frac{\sqrt l}{8(l + 1)\left( l + \frac{1}{2} \right)} ~.
\ee
Expression (B.17) as compaired with the $\Delta \varepsilon^{D=0}$
gives the correction   to $\Delta \varepsilon$ of order of $\sim 10\%$.
The dispersion correction in the slope occures even smaller (to the analogy
with the ratio of quark mass  contribution (105) in the energy and in the
Regge slope (104), discussed in the main text).
In conclusion of this Appendix we note that relative
contribution of the dispersion from higher powers of $\mu$ and $\nu$
to the energy is  also either
of the same order or even smaller.

\newpage

\setcounter{equation}{0}
\renewcommand{\theequation}{C.\arabic{equation}}

{\bf Appendix C}
\vspace{5mm}

In this Appendix we derive  expression (87) for
Hamiltonian $H^{(1)}_r$ which describes radial
dynamics in the limit of large $l$ and calculate the corrections to the
spectrum of pure string zero approximation (85). We exploit the method,
developed in [1] and consider in a unif\/ied way two limiting cases of the
full straight-line string with $m_1 = m_2 = m,~ \mu_1 = \mu_2 = \mu~
(K = 2)$ and the half of the string with $m_1 = m,~ m_2 \to \infty~
(K = 1)$.
We start here with Hamiltonian (79) of the main text. Our strategy will
be to expand (in two limiting cases under consideration)
the following part of (79)
\be
\begin{array}{ll}
  2 \left( \frac{l (l + 1) \int \frac{\sigma^2 d \beta}{\nu}}
  {\mu_1 (1 - \zeta)^2 + \mu_2 \zeta^2 + \int d \beta \nu (\beta - \zeta)^2}
   \right)^{1/2} =
  2 \left( \frac{l (l + 1) \int \frac{\sigma^2 d \beta}{\nu}}
  {\frac{\mu}{K} + \int d \beta (\beta - \frac{K - 1}{2})^2 \nu}
   \right)^{1/2}  \approx  \\
\approx 2 \left( \frac{ \sigma^2 l (l + 1) \int \frac{ d \beta}{\nu}}
  { \int d \beta  (\beta - \frac{K - 1}{2})^2 \nu}
   \right)^{1/2} +   (-1) \frac{\mu}{K}
   \frac{\left( \sigma^2 l (l + 1) \int \frac{ d \beta}{\nu} \right)^{1/2}}
   {\left( \int d \beta  (\beta - \frac{K - 1}{2})^2 \nu \right)^{3/2} }
   + \frac{3}{4} \left( \frac{\mu}{K} \right)^2
   \frac{ \left( \sigma^2  l (l + 1) \int \frac{ d \beta}{\nu} \right)^{1/2}}
   {\left( \int d \beta  \left( \beta - \frac{K - 1}{2} \right)^2 \nu
   \right)^{5/2} }  \\
\end{array}
\ee
and then integrate this approximate expression over $\mu$ to end up with
Hamiltonian (87).

After expansion (C.1)
 the terms in eq.(79) independent of $\mu_i$ are collected into the pure
string Hamiltonian (83) which gives zero approximation to the problem.
To obtain the leading radial corrections one is to find the extremal value
(84) for $\nu^{(0)}$ from eq.(83) (in the way similar to that of
[12, 1]) and substitute it into $H$ with the result
$$   H ~ = ~ ( K \pi \sigma \sqrt{l (l + 1)} )^{1/2}  + \frac{1}{2}
  \left[ \frac{\vec p^{~2}_r + m^2}{\mu/2} + \frac{3 K^2 \mu^2}
  {2(K \pi \sigma l)^{1/2}} +  \right.  $$
\be
  \left. + \frac{(\pi \sigma)^2 (r - r_0)^2}{2(K \pi \sigma l)^{1/2}}
  \right] ~ \equiv ~ \varepsilon^{(0)}_l + H^{(1)}_r
\ee
where the first term corresponds to the pure string result (85).
Introducing new variables
\be
   \tilde{\mu} = \sqrt{K}\mu~, ~~~~ \tilde p = \sqrt{K} p~, ~~~~
   \tilde r - \tilde r_0 = \frac{r - r_0}{\sqrt{K}}
\ee
one obtains the unified radial Hamiltonian $H^{(1)}_r = H -
\varepsilon^{(0)}_l$ for h.l. and l.l. mesons
\be
   H_r = \frac{1}{2} \sqrt{K} \left[ \frac{\tilde p^2 + \sum^K_{i=1} m_i^2}
   {\tilde{\mu} } + \frac{3 \tilde{\mu}^2}{2(\pi \sigma l)^{1/2}} +
   \frac{(\pi \sigma)^2 (\tilde r - \tilde r_0)^2}{2(\pi \sigma l)^{1/2}}
   \right] \equiv \sqrt{K} \tilde H_r
\ee
so that $H = \varepsilon_i^{(0)} + H^{(1)}_r \equiv \sqrt K (E^{(0)}
_l + \tilde H^{(1)}_r)$.
After the integration over $\tilde{\mu}$, which amounts to the substitution of
its extremal value
\be
   \tilde{\mu} ~ = ~ \left[ \frac{\left( \tilde p^2 + \sum^K_{i=1} m^2_i
\right)}
   {3} (\pi \sigma l)^{1/2} \right]^{1/3}
\ee
we arrive at the effective radial Hamiltonian
\be
   H^{(1)}_r ~ = ~\frac{1}{2} \sqrt{K} \left[ \frac{3^{4/3}}{2} \cdot
   \frac{\left( \tilde p^2 + \sum^K_{i=1} m^2_i \right)^{2/3}}
   {(\pi \sigma l)^{1/6}} + \frac{(\pi \sigma)^2 (\tilde r - \tilde r_0)^2}
   {2 (\pi \sigma l )^{1/2} } \right]
\ee
Introducing instead of $(\tilde r - \tilde r_0)$ a new dimensionless
variable $x$
\be
   \frac{(\tilde r - \tilde r_0)}{x} ~ = ~ \frac{p_x}{\tilde p_r} ~ = ~
   \frac{3^{2/5} \cdot l^{1/10} }{(\pi \sigma)^{1/2}}
\ee
one can represent $H^{(1)}_r$ in the following form
\be
   H^{(1)}_r = \sqrt{K} \left( \frac{3^{4/5} (\pi \sigma)^{1/2}}{4}
   l^{-3/10} \right) \left[ \left( p^2_x + 3^{4/5} l^{1/5}
   \frac{\sum^K_{i=1} m^2_i}{\pi \sigma} \right)^{2/3} + x^2
   \right]
\ee
which is convenient for numerical calculations.

In order to obtain an approximation to the spectrum $\varepsilon^{(1)}
_{n_r, l}$ of $H_r^{(1)}$  we consider (C.4) for the restricted class of
functions $\tilde{\mu} $ independent on $\tau$. This procedure usually
gives the accuracy about 5\% - 10\% for the states with not very large
$n$ (see also the discussion in Appendix B)
[1], [3]. To this end [7] we first evaluate $\varepsilon^{(1)}_{n_r, l}$
for a given $\tilde{\mu}$
\be
\begin{array}{r}
   \varepsilon^{(1)}_{n_r, l} (\tilde{\mu}) = \sqrt{K} \left[ \frac{(\pi
\sigma)}
   {\left( 2 (\pi \sigma l )^{1/2} \right)^{1/2}} \cdot \frac{1}
   {\tilde{\mu}^{1/2}} \left( n_r + \frac{1}{2} \right) +
   \frac{\sum^K_{i=1} m^2_i}{2\tilde{\mu}} + \frac{3}{4}
   \frac{\tilde{\mu}^2}{(\pi \sigma l)^{1/2}} \right] \equiv    \\
   = \sqrt{K} E^{(1)}_{n_r, l} (\tilde{\mu})
\end{array}
\ee
The integration over $\tilde{\mu}$, as well as in the general case,
amounts to the substitution of its extremal value $\frac{d \varepsilon^{(1)}}
{d \tilde{\mu}} \left|_{\tilde{\mu}_{ext}} = 0 \right.$ in (C.9) .
The extremal condition has the form
\be
   \frac{(\pi \sigma)}{\left( 2 (\pi \sigma l)^{1/2} \right)^{1/2}}
   \cdot \frac{\left( n_r + \frac{1}{2} \right)}{\tilde{\mu}^{3/2}} +
   \frac{\sum^K_{i=1} m^2_i}{\tilde{\mu}^2} = 3 \frac{\tilde{\mu}}
   {(\pi \sigma l)^{1/2}}
\ee

and is dif\/ficult to solve analitically.
Therefore we find the solution in two dif\/ferent asymptotics. The first one
corresponds to the relativistic radial dynamics
\be
   <~ p^2_r ~>  ~~\gg~~ ~\sum^K_{i=1} m^2_i
\ee
that in the leading order gives
\be
   \tilde{\mu} ~ = ~ \left[ \frac{(\pi \sigma)^{5/4} l^{1/4}}{3\sqrt{2}}
   \left( n_r + \frac{1}{2}\right) \right]^{2/5}
\ee
and therefore the spectrum (C.9) can be represented in the form
$$    E^{(1)}_{n_r, l} = E^{(1,0)}_{n_r, l} + E^{(1,1)}_{n_r, l} $$
with
\be
\begin{array}{ll}
    E^{(1,0)}_{n_r, l} = \frac{5}{4} \cdot \left( \frac{3}{4} \right)^{1/5}
    (\pi \sigma)^{1/2} l^{-3/10} \left( n_r + \frac{1}{2} \right)^{4/5} \\
    E^{(1,1)}_{n_r, l} = \left( \frac{3}{4} \right)^{2/5}
    \frac{\sum^K_{i=1} m^2_i}
    {(\pi \sigma)^{1/2}} l^{-1/10} \left( n_r + \frac{1}{2} \right)^{-2/5}
\end{array}
\ee

We note that $E^{(0)}_{n_r, l}$ is the radial correction corresponding to the
massless quark case (84) which was evaluated in [1].

To improve the accuracy we take into account that usually the value of
$\varepsilon $ obtained according to the time-independent
$\mu $ ansatz (C.9), (C.10) is $1,1$ times larger than the exact one
(see [3] for a comparison). Therefore one is to multiply eqs.(C.13)
additionally by a factor $\approx 0.9$.

The condition (C.11) can be reformulated as $E^{(1,1)}_{n_r, l}
/E^{(1,0)} _{n_r, l} \ll 1$ with the result
\be
   \left( \frac{\sum^K_{i=1} m^2_i}{\pi \sigma}  \right)^5
   ~~~\ll~~~ \frac{\left( n_r +
    \frac{1}{2} \right)^6}{l}
\ee

Together with the condition, that zero order  contribution (80),
$E^{(0)}_{str} = \sqrt{\pi \sigma (l(l+1))^{1/2}}$, is much more than
$E^{(1,0)}_{n_r, l}$  of eq.(C.13)
\be
   l / (n_r + 1/2)  ~~\gg~~ 1
\ee
one obtains the domain of quantum numbers
where  expression for the spectrum (C.13) is valid.

In the opposite nonrelativistic asymptotics
for radial dynamics
\be
   < p^2_r > ~~\ll~~ \sum^K_{i=1} m^2_i
\ee
one obtains in the leading order
\be
   \tilde{\mu} ~ = ~ \left[ \frac{\sum^K_{i=1} m^2_i (\pi \sigma l)^{1/2}}
   {3} \right]^{1/3}
\ee
which gives rise to
\be
\begin{array}{ll}
   E^{(1,0)}_{n_r, l} = \frac{3^{4/3}}{4} \cdot \frac{(\sum_i m^2_i)^{2/3}}
   {(\pi \sigma)^{1/6}}  l^{-1/6}   \\
   E^{(1,1)}_{n_r, l} ~ = ~ \left(\frac{3}{8} \right)^{1/6} \cdot \frac{(\pi
   \sigma)^{2/3}} {(\sum_i m^2_i)^{1/6} }  l^{-1/3} \left( n_r + \frac{1}{2}
   \right)  ~.
\end{array}
\ee

As well as in the previous case the requirements of selfconsistency,
$E^{(1,0)}_{n_r, l} \gg E^{(1,1)}_{n_r, l},~
E^{(0)}_{str} \gg E^{(0)}_{n_r, l}$, ~determine the conditions
\be
   \left( \frac{\sum^K_{i=1} m^2_i}{\pi \sigma} \right)^5 ~~\gg~~
   \frac{(n_r + 1/2)^6}{l}
\ee
and
\be
   l ~~\gg~~ \left(  \sum^K_{i=1} m^2_i /\pi \sigma \right)
\ee
respectively.

To summarize we calculated  radial corrections in different asymptotics of
the string
regime which is valid under conditions (C.15), (C.20) that
can be expressed in the unified way
\be
 \frac{l}{n_r + \frac{\sum_i m^2_i}{\pi \sigma} + \frac{1}{2}} ~~ \gg ~~ 1 ~ .
\ee


\newpage
\centerline{\bf Figure Captions}

\vspace{10mm}

\begin{tabular}{p{15mm}p{130mm}}

Fig. 1. & Regge trajectories of heavy-light and light-light mesons (with
$m_q = 0$ for light quarks). Crosses are the results of the approximation of
eqs.(64) - (66) and open circles are calculated with an account of the
leading correction, eqs. (69) - (71) (they are not shown for $l = 5$
since they practically coincide with open squares).
Open squares are the predictions of
the large $l$ approximation (91) with an account of the correction (89).
Black dots are the results of the numerical calculation of ref. [5]

\end{tabular}

\end{document}